\documentclass{PoS}

\title{Inclusive Deep Inelastic Scattering at High $Q^2$ with Longitudinally Polarised Lepton Beams at HERA}

\ShortTitle{Inclusive DIS at High $Q^2$ with Longitudinally Polarised Lepton Beams at HERA}

\author{\speaker{Zhiqing ZHANG}\thanks{On behalf of the H1 Collaboration.}\\
        Laboratoire de l'Acc\'el\'erateur Lin\'eaire, Univ.\ Paris-Sud and IN2P3/CNRS, Orsay, France\\
        E-mail: \email{zhang@lal.in2p3.fr}}

\abstract{Inclusive $e^\pm p$ single and double differential cross sections for neutral (NC) and charged current (CC) deep inelastic scattering processes are measured with the H1 detector at HERA. The data were taken at a centre-of-mass energy of $\sqrt{s}=319$\,GeV with a total integrated luminosity of 327.8\,pb$^{-1}$ shared between two lepton beam charges and two longitudinal lepton polarisation modes. The differential cross sections are measured in the range of negative four-momentum transfer squared, $Q^2$, between 60 and 50\,000\,GeV$^2$, and Bjorken $x$ between 0.0008 and 0.65. The measurements are combined with earlier published unpolarised H1 data to improve statistical precision and used to determine the structure function $xF_3^{\gamma Z}$. A measurement of the structure function $F_2^{\gamma Z}$, sensitive to parity violating effects in NC, is presented for the first time. The polarisation dependence of the CC total cross section is also measured. The new measurements are well described by a next-to-leading order QCD fit based on all published H1 inclusive cross section data which are used to extract the parton distribution functions of the proton.}

\FullConference{The European Physical Society Conference on High Energy Physics -EPS-HEP2013\\
		18-24 July 2013\\
		Stockholm, Sweden}

\begin{document}

\section{Introduction}
The $ep$ collider HERA used to be the largest electron microscope of the world. 
The inclusive neutral current (NC) and charged current (CC) deep inelastic scattering (DIS) cross section data measured by H1 and ZEUS at HERA\,I (1992-2000)~\cite{h1zeus-hera1}, where the unpolarised electron\footnote{In this proceedings, 
``electron" refers generically to both electrons and positrons. Where distinction is required the terms $e^-$ and $e^+$ are used.} beam of up to 27.5\,GeV collided with the unpolarised proton beam of up to 920\,GeV, have been the primary source for constraining of parton distribution functions (PDFs) of the proton. In 
HERA\,II (2003-2007), spin rotators were installed around the interaction points 
of H1 and ZEUS to provide longitudinally polarised electron beam there. Most of 
the data at HERA\,II were taken at the nominal proton beam energy $E_p$ of 920\,G
eV in four distinct data sets: $e_R^-p$, $e_L^-p$, $e_R^+p$ and $e_R^+p$, i.e.\ 
right- and left-handed polarised electron and positron beams scattering with proton beam. The corresponding integrated luminosity (mean polarisation $P_e$) values\footnote{The luminosity values given here differ from~\cite{h1-hera2} by -1.8\% due to the revised measurement based on the QED Compton events~\cite{h1qed}.}  are 46.5\,pb$^{-1}$ ($+36.0$\%), 102.6\,pb$^{-1}$ ($-25.8$\%), 99.5\,pb$^{
-1}$ ($+32.5$\%) and 79.3\,pb$^{-1}$ ($-37.0$\%), respectively. The $e^-p$ data 
corresponds to an almost tenfold increase in luminosity over the HERA\,I data set. 
%Towards the end of data taking, two data sets with lower proton beam energies 
%of 575\,GeV and 460\,GeV were taken for a direct measurement of longitudinal str
%ucture function $F_L$ (Sec.~\ref{sec:fl})~\cite{fl}.

The new results reported here and published in~\cite{h1-hera2} concern the final NC and CC cross sections at high $Q^2$ measured using all HEAR-II data and a new QCD analysis of all H1 data to obtain a new set of PDFs, H1PDF\,2012.
 
To appreciate the constraint on PDFs from the inclusive NC and CC cross sections, it is helpful to write down explicitly the expressions of the differential NC and CC cross sections in terms of different structure functions  
\begin{eqnarray}
\frac{d^2\sigma^\pm_{\rm NC}}{dxdQ^2}&=&\frac{2\pi\alpha^2}{xQ^4}\left(Y_+\tilde{F}^\pm_2\mp Y_-x\tilde{F}_3^\pm -y^2\tilde{F}_L^\pm\right)\,,\label{eq:nc2}\\
\frac{d^2\sigma^\pm_{\rm CC}}{dxdQ^2}&=&(1\pm P_e)\frac{G^2_F}{2\pi x}\left[\frac{M^2_W}{M^2_W+Q^2}\right]^2\left(Y_+W_2^\pm\mp Y_-xW_3^\pm-y^2W_L^\pm\right)\,,\label{eq:cc2}
\end{eqnarray}
where $Y_\pm=1\pm(1-y)^2$ and $y$ characterises the inelasticity of the interaction and is related to Bjorken $x$, $Q^2$ and centre-of-mass energy $\sqrt{s}$ by $y=Q^2/(xs)$, $\alpha\equiv\alpha(Q^2=0)$ is the fine structure constant, $G_F$ is the Fermi constant and $M_W$ is the propagator mass of the $W$ boson. For the NC interaction, the generalised structure functions $\tilde{F}_2^\pm$ and $x\tilde{F}_3^\pm$ can be further decomposed as
\begin{eqnarray}
\tilde{F}_2^\pm&=&F_2-(v_e\pm P_ea_e)\kappa\frac{Q^2}{Q^2+M^2_Z}F_2^{\gamma Z}+(a^2_e+v^2_e\pm P_e2v_ea_e)\kappa^2\left[\frac{Q^2}{Q^2+M^2_Z}\right]^2F_2^Z\,,\label{eq:f2}\\
x\tilde{F}_3^\pm&=&-(a_e\pm P_ev_e)\kappa\frac{Q^2}{Q^2+M^2_Z}xF_3^{\gamma Z}+(2a_ev_e\pm P_e[v^2_e+a^2_e])\kappa^2\left[\frac{Q^2}{Q^2+M^2_Z}\right]^2xF_3^Z\,,\label{eq:f3}
\end{eqnarray}
with $M_Z$ being the mass of the $Z$ boson, $\kappa^{-1}=4\frac{M^2_W}{M^2_Z}\left(1-\frac{M^2_W}{M^2_Z}\right)$ in the on-mass-shell scheme, $v_e$ and $a_e$ the vector and axial-vector couplings of the electron to the $Z$ boson. In Eq.(\ref{eq:f2}), $F_2$ corresponds to the dominant electromagnetic structure function of photon exchange, the other structure function terms are due to photon-$Z$ interference and pure $Z$ exchange. The generalised longitudinal structure function $\tilde{F}_L$ may be similarly decomposed. 

In the quark-parton model (QPM), the structure functions $F_2$, $F_2^{\gamma Z}$ and $F_2^Z$ are related to the sum of the quark and anti-quark densities $\left[F_2, F_2^{\gamma Z}, F_2^Z\right]=x\sum_q\left[e^2_q, 2e_qv_q, v^2_q+a^2_q\right]\{q+\bar{q}\}$ ($a_q$, $v_q$ and $a_q$ being the electric charge of quark $q$ and its vector and axial-vector couplings to the $Z$ boson) and the structure functions $xF_3^{\gamma Z}$ and $xF_3^Z$ to the difference between quark and anti-quark densities $\left[xF_3^{\gamma Z}, xF_3^Z\right]=x\sum_q\left[ 2e_qa_q, 2v_qa_q\right]\{q-\bar{q}\}$. The longitudinal structure function $F_L$ is zero in QPM because of helicity conservation. In QCD, the gluon emission gives rise to a non-vanishing $F_L$. 
%Measuring the structure function $F_L$ therefore provides a way of studying the gluon density and a test of perturbative QCD.

\section{Double differential NC and CC cross sections}
\begin{figure}[htb]
\begin{center}
\includegraphics[width=.495\textwidth]{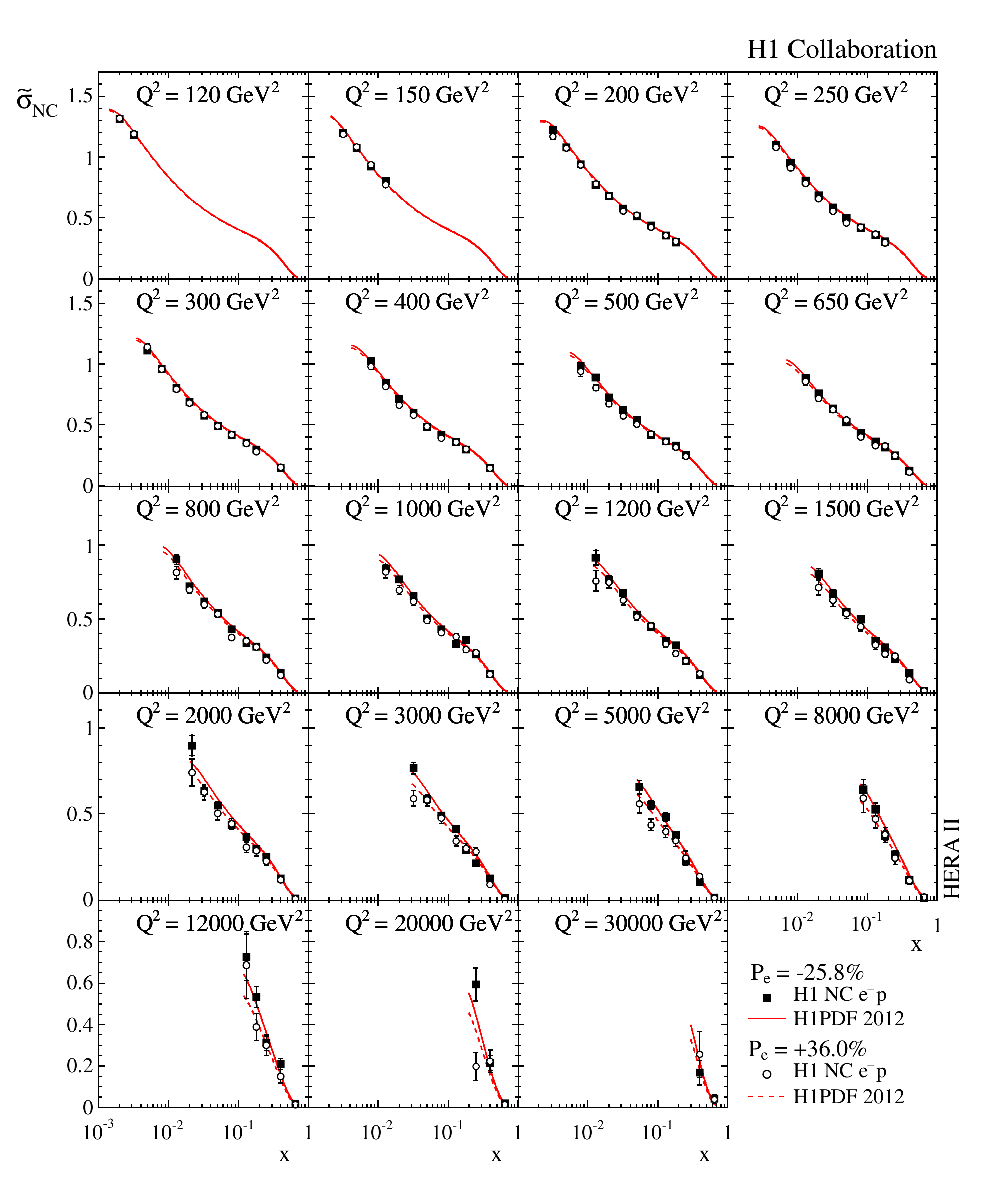}
\includegraphics[width=.495\textwidth]{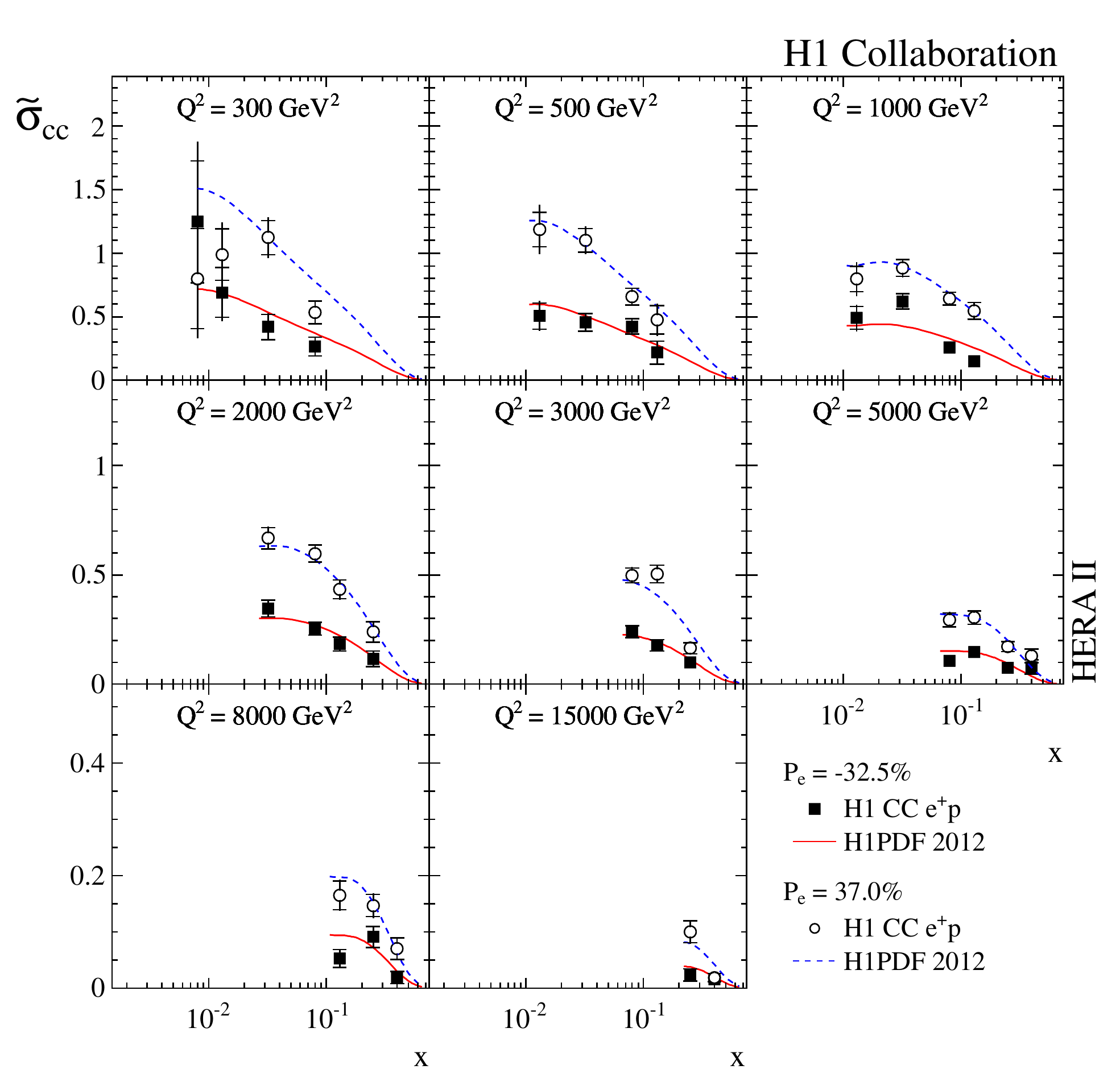}
\end{center}
\vspace{-7mm}
\caption{NC and CC reduced cross sections $\tilde{\sigma}_{\rm NC}$ (left) and $\tilde{\sigma}_{\rm CC}$ (right) for $e^- p$ and $e^+ p$ data sets, respectively. The inner error bars represent the statistical uncertainties and the full error bars the total uncertainties. The curves show the corresponding expectations from H1PDF\,2012.}
\label{fig:sig2}
\end{figure}

Double differential NC cross sections $d^2\sigma_{\rm NC}/dxdQ^2$ have been measured for the four data sets corresponding to the left and right handed polarised lepton beam $e^\pm$ in the kinematic region $120\leq Q^2\leq 50\,000\,{\rm GeV}^2$ and $0.002\leq x\leq 0.65$.  The cross section results for the $e^- p$ data are shown in Fig.~\ref{fig:sig2} (left) in a reduced form ($\tilde{\sigma}_{\rm NC}$) by removing the kinematic factor $2\pi\alpha^2 Y_+/xQ^4$ in Eq.(\ref{eq:nc2}). The cross sections for the left and right handed polarisation states are found to agree at low $Q^2 (\lesssim 1\,000\,{\rm GeV}^2$). At higher $Q^2$ and lower $x$, deviations are observed between the two as expected from the parity violation of $Z$ boson exchange. Given the small sensitivity to the lepton beam polarisation, at high $y$ region covering $0.19<y<0.63$, $60\leq Q^2\leq 800\,{\rm GeV}^2$ and $0.0008\leq x\leq 0.0105$ additional NC cross sections have been measured combing the left and right handed polarisation data sets of the $e^\pm p$ interactions.

Double differential CC cross section $d^2\sigma_{\rm CC}/dxdQ^2$ have also been measured for the same data sets in the kinematic region $300\leq Q^2\leq 30\,000\,{\rm GeV}^2$ and $0.008\leq x\leq 0.4$. One example is shown in Fig.~\ref{fig:sig2} (right) also in a reduced form ($\tilde{\sigma}_{\rm CC}$) by removing the kinematic factor $G^2_F/2\pi x \left[M^2_W/(M^2_W+Q^2)\right]^2$ in Eq.(\ref{eq:cc2}). The CC reduced cross sections for the $L$ and $R$ data sets are very different for all $Q^2$ as parity violation is maximal with $W$ boson exchange. These cross sections agree well with the expectation from H1PDF\,2012 fit. Both the statistical and systematic precision have substantially improved with respect to the corresponding measurements from HERA\,I with the unpolarised lepton beams.

\section{NC polarisation asymmetry and structure functions $F_2^{\gamma Z}$ and $xF_3^{\gamma Z}$}
\begin{figure}[htb]
\begin{center}
\includegraphics[width=.495\textwidth]{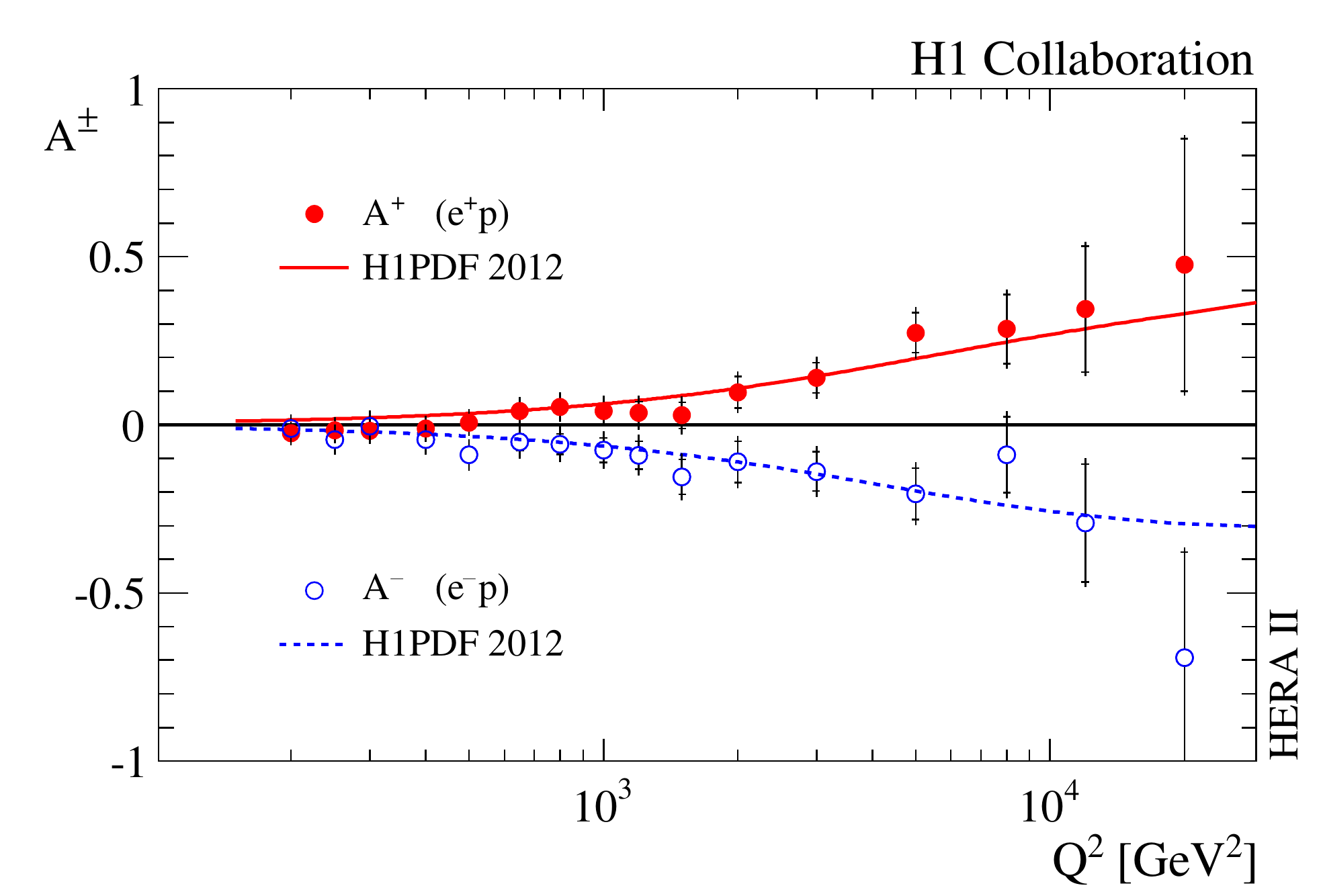}\\
\includegraphics[width=.495\textwidth]{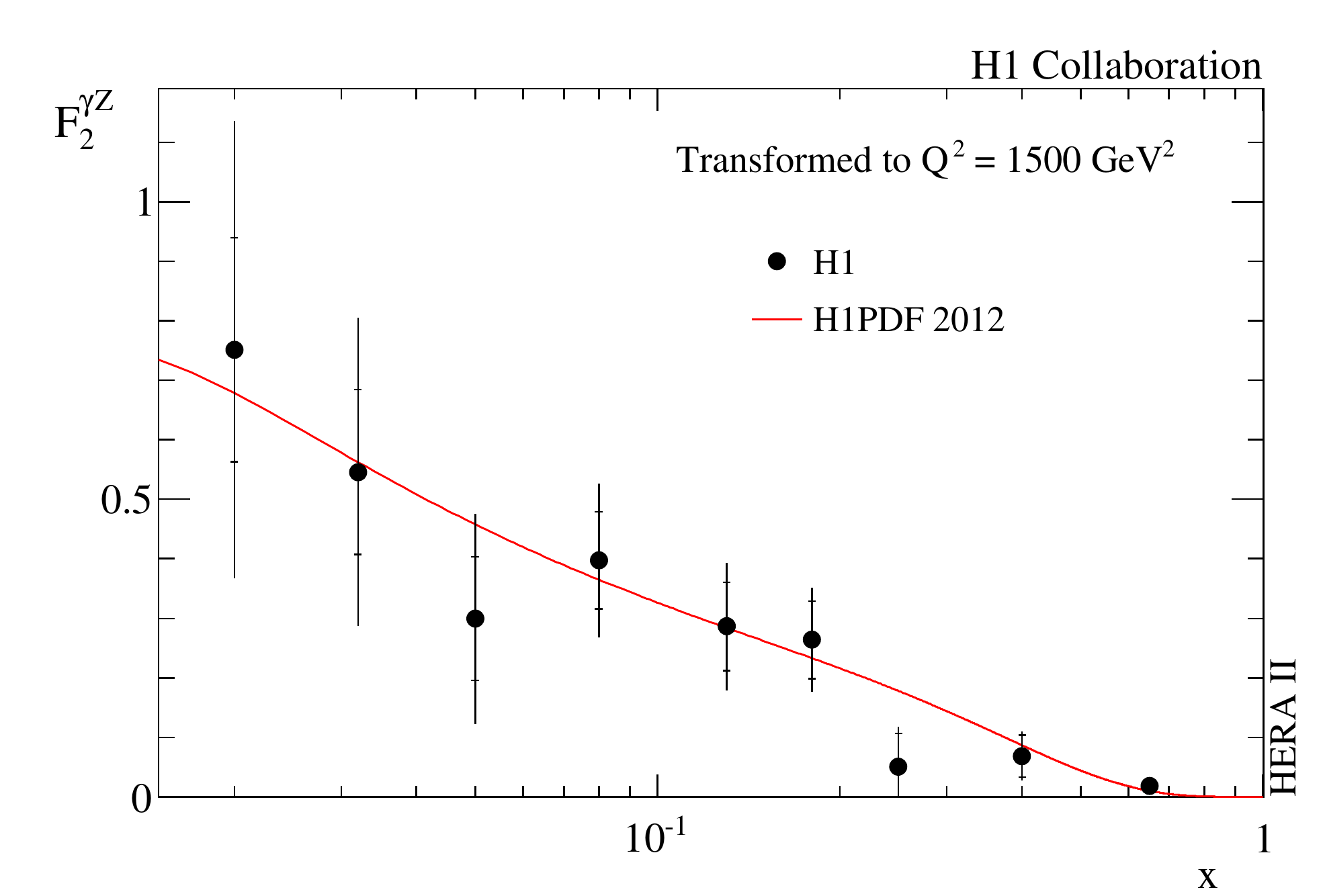}
\includegraphics[width=.495\textwidth]{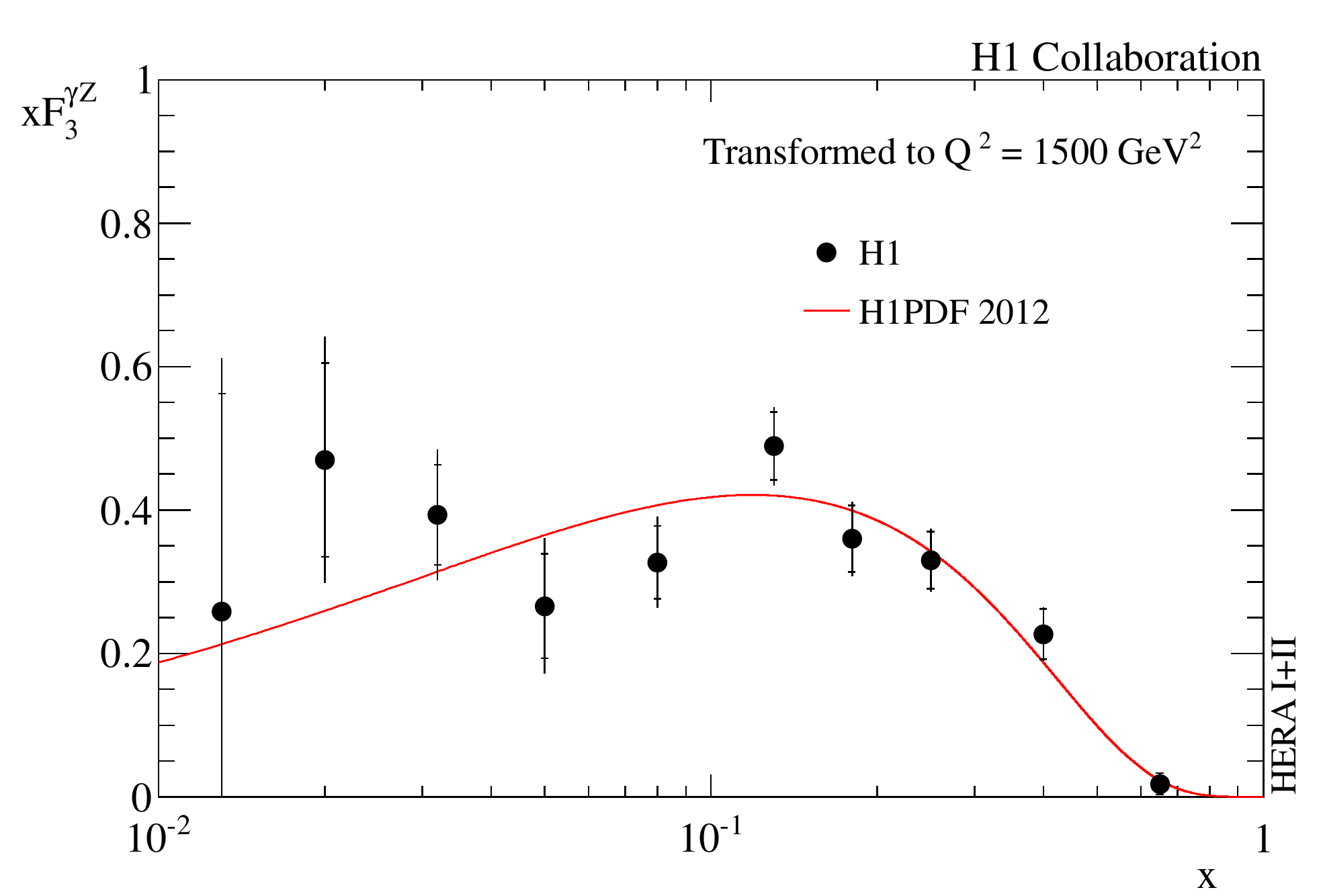}
\end{center}
\vspace{-7mm}
\caption{$Q^2$ dependence of the polarisation asymmetry $A^\pm$ (top) and structure functions $F_2^{\gamma Z}$ (bottom-left) and $xF_3^{\gamma Z}$ (bottom-right) transformed to $Q^2=1\,500\,{\rm GeV}^2$ for data (points) and the expectations from H1PDF\,2012 (curves). The inner error bars represent the statistical uncertainties and the full error bars correspond to the total measurement uncertainties.}
\label{fig:pol_sfs}
\end{figure}

The Standard Model (SM) predicts a difference in the NC cross section for leptons with different helicity states arising from the chiral structure of the neutral electroweak exchange. With longitudinally polarised lepton beams in HERA\,II such electroweak effects can be tested through a polarisation asymmetry, $A^\pm$, defined as
\begin{equation}
A^\pm=\frac{2}{P^\pm_L-P^\pm_R}\cdot\frac{\sigma^\pm (P^\pm_L)-\sigma^\pm (P^\pm_R)}{\sigma^\pm (P^\pm_L)+\sigma^\pm (P^\pm_R)}\,,
\end{equation}
where $P^\pm_L$ and $P^\pm_R$ are the longitudinal lepton beam polarisation in the $e^\pm p$ $R$ and $L$ data sets. The measured asymmetry is shown in Fig.~\ref{fig:pol_sfs} (top). 
The magnitude of the asymmetry is observed to increase with increasing $Q^2$ and is positive in $e^+p$ and negative in $e^-p$ scattering. The data are in good agreement with the SM using H1PDF\,2012 and confirm the parity violation effects of the electroweak interactions at large $Q^2$.

From the measured NC cross sections and using Eqs.(\ref{eq:nc2}), (\ref{eq:f2}) and (\ref{eq:f3}), one can extract the structure function $F_2^{\gamma Z}$. The measurement is performed for $Q^2\geq 200\,{\rm GeV}^2$. Only a weak $Q^2$ dependence is expected and therefore the measurement is transformed to a common $Q^2$ value of 1\,500\,GeV$^2$ using the H1PDF\,2012 fit and is averaged in each $x$ bin. The result is shown in Fig.~\ref{fig:pol_sfs} (bottom-left) in comparison to the H1PDF\,2012 fit. This is the first measurement of the structure function $F_2^{\gamma Z}$.

The structure function $xF_3^{\gamma Z}$, probing the valence quarks of the proton, has been determined previously by H1 with limited statistical precision due in particular to the small integrated luminosity available in $e^- p$ interactions at HERA\,I~\cite{hera1-em,hera1}. Combining the new HERA\,II data with that of HERA\,I, the structure function has been updated in Fig.~\ref{fig:pol_sfs} (bottom-right) with two new data points at low $x$ and substantially improved precision for other data points.

\section{Total CC cross section $\sigma_{\rm CC}^{\rm tot}$ and NC and CC cross section $d\sigma_{\rm NC, CC}/dQ^2$}
\begin{figure}[htb]
\begin{center}
\includegraphics[width=.495\textwidth]{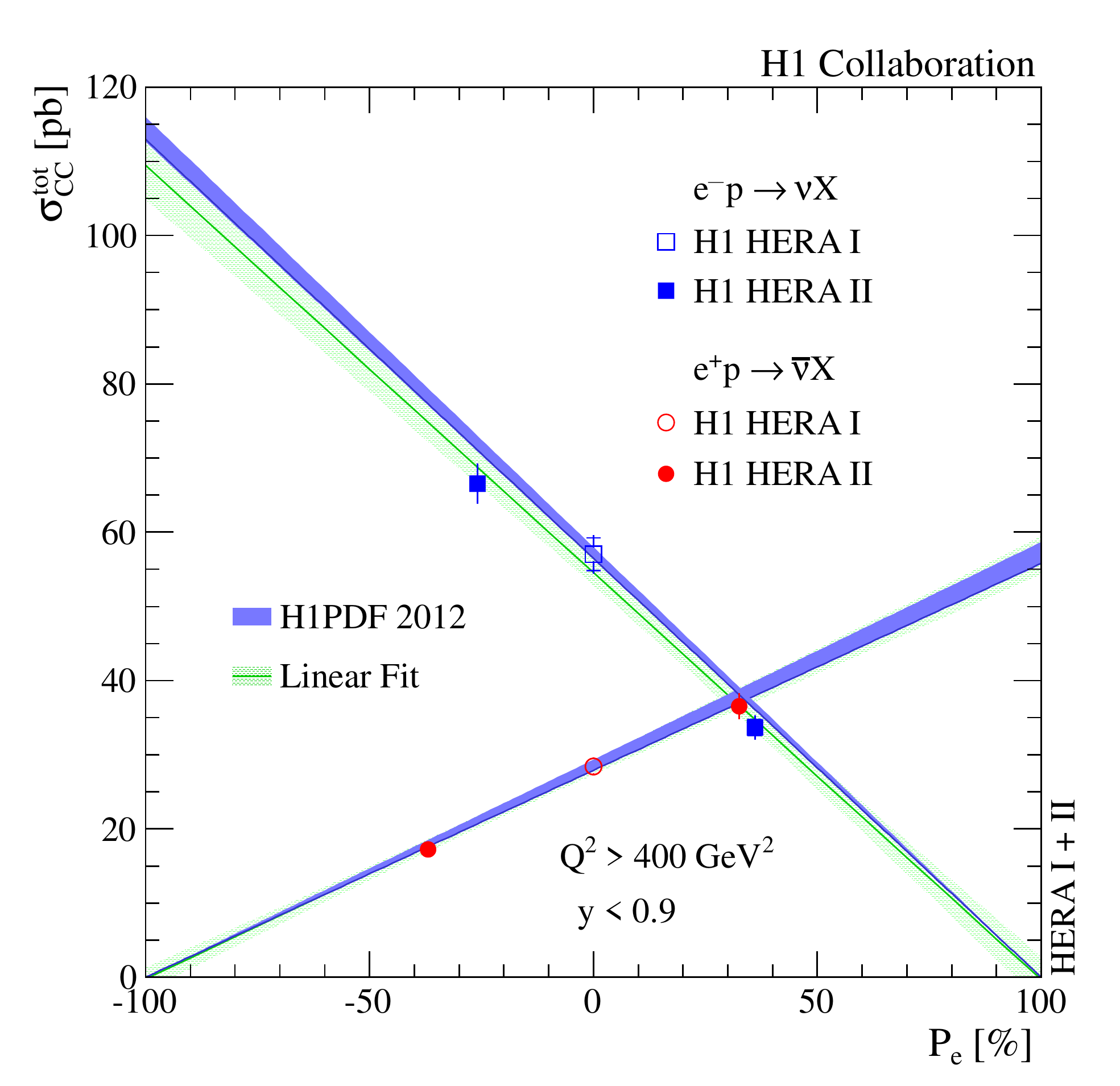}
\includegraphics[width=.495\textwidth]{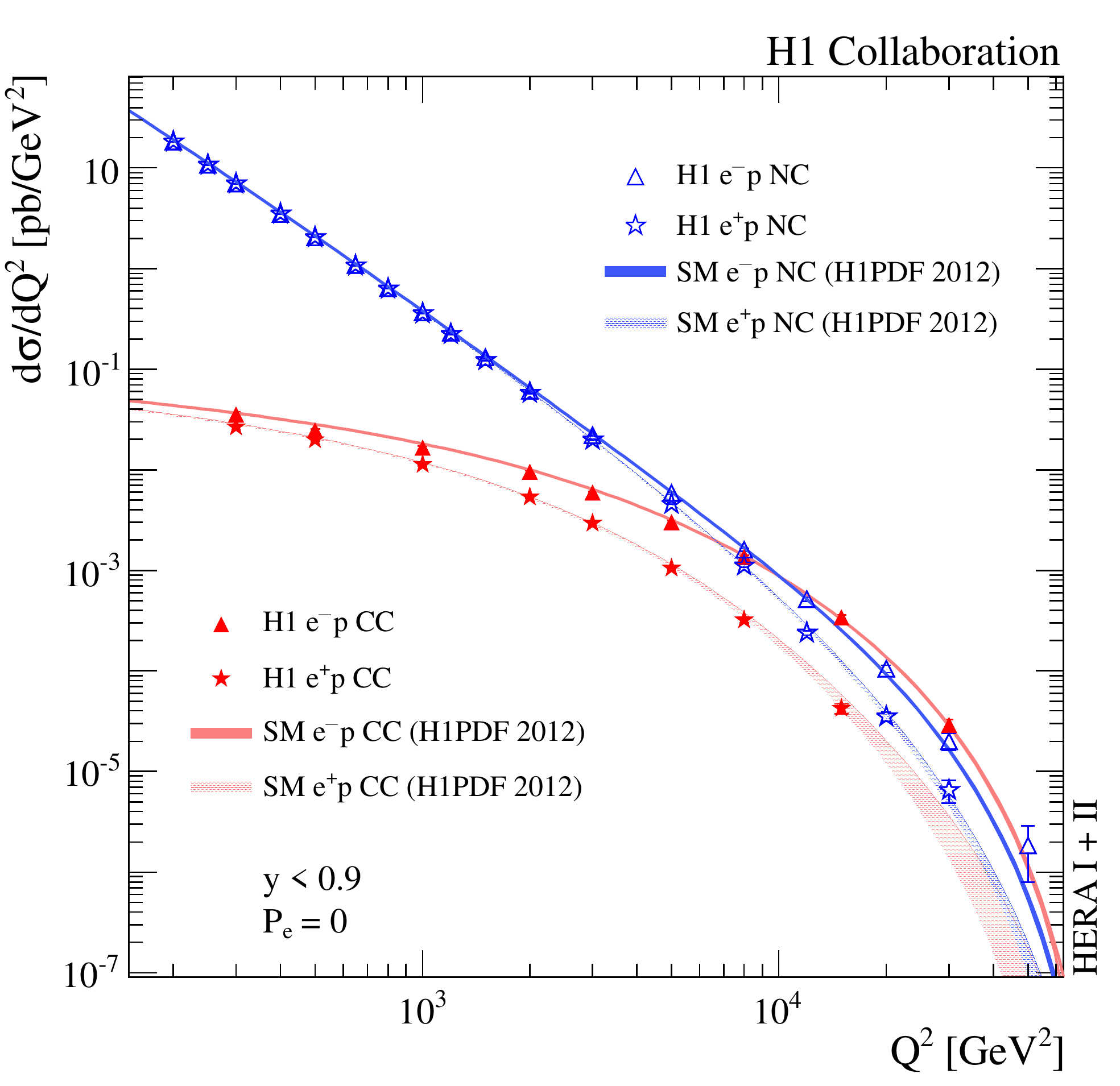}
\end{center}
\vspace{-7mm}
\caption{\underline{Left}: Dependence of the $e^\pm p$ CC cross sections on the longitudinal lepton beam polarization $P_e$. \underline{Right}: $Q^2$ dependence of the NC and CC cross sections $d\sigma/ dQ^2$ for the combined HERA\,I+II unpolarised $e^\pm p$ data. In both figures, the inner and full error bars represent the statistical and total errors, respectively.}
\label{fig:cctot-nccc}
\end{figure}

The total CC cross sections $\sigma^{\rm tot}_{\rm CC}$ for $e_{L, R}^\pm p$ scattering have been  measured in the kinematical region of $Q^2>400\,{\rm GeV}^2$ and $y<0.9$. Together with the corresponding cross sections measured at HERA\,I with unpolarised $e^\pm$ beams, the linear dependence of the cross sections on $P_e$ (Fig.\,\ref{fig:cctot-nccc}(left)) can be verified for both $e^-p$ and $e^+p$ interactions (in the latter case with more data than that previously published~\cite{cctot05}). The data are consistent with the absence of right handed weak charged currents. 

The single differential NC and CC cross sections $d\sigma_{\rm NC,CC}/dQ^2$ have been measured for $y<0.9$ and in $Q^2$ varying over more than two orders of magnitude (the combined version with unpolarised HERA\,I $e^\pm p$ data is shown in Fig.\,\ref{fig:cctot-nccc}(right)). The NC cross sections exceed the CC cross sections at $Q^2\simeq 200\,{\rm GeV}^2$ by more than two orders of magnitude. The steep decrease of the NC cross section with increasing $Q^2$ is due to the dominating
 photon exchange cross section which is proportional to $1/Q^4$. In contrast the
 CC cross section is proportional to $[M_W^2 /(Q^2 + M_W^2 )]^2$ and approaches 
a constant value at $Q^2 \simeq 300\,{\rm GeV}^2$. The NC and CC cross sections 
are of comparable size at $Q^2 \sim 10^4\,{\rm GeV}^2$, where the photon and $Z$
 exchange contributions to the NC process are of similar size to those of $W^\pm$ exchange to the CC process. These measurements thus illustrate the unified behavior of the electromagnetic and the weak interactions in DIS.

\section{H1PDF\,2012}
\begin{figure}[htb]
\begin{center}
\includegraphics[width=.4\textwidth]{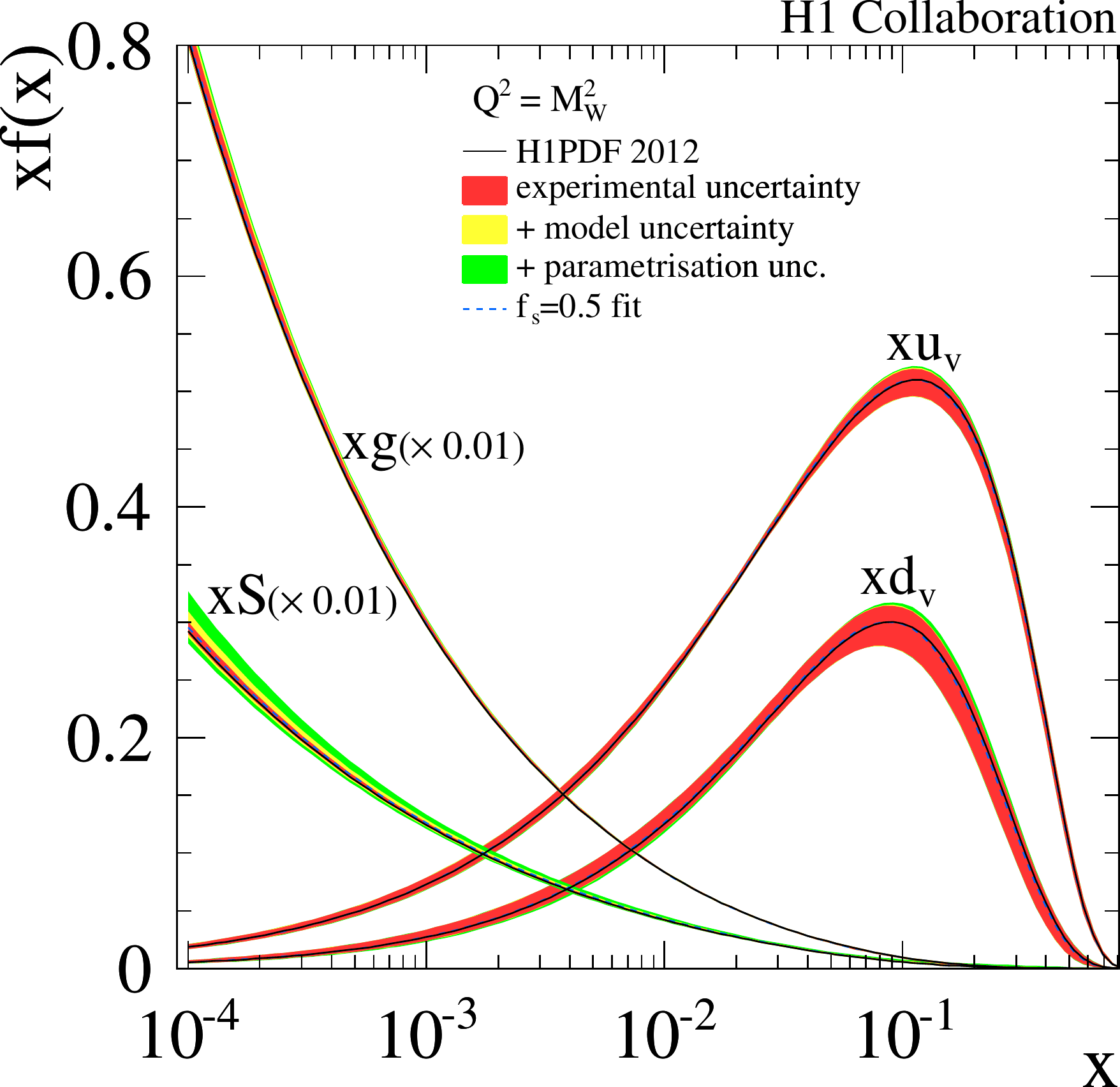}\hspace{2mm}
\includegraphics[width=.58\textwidth]{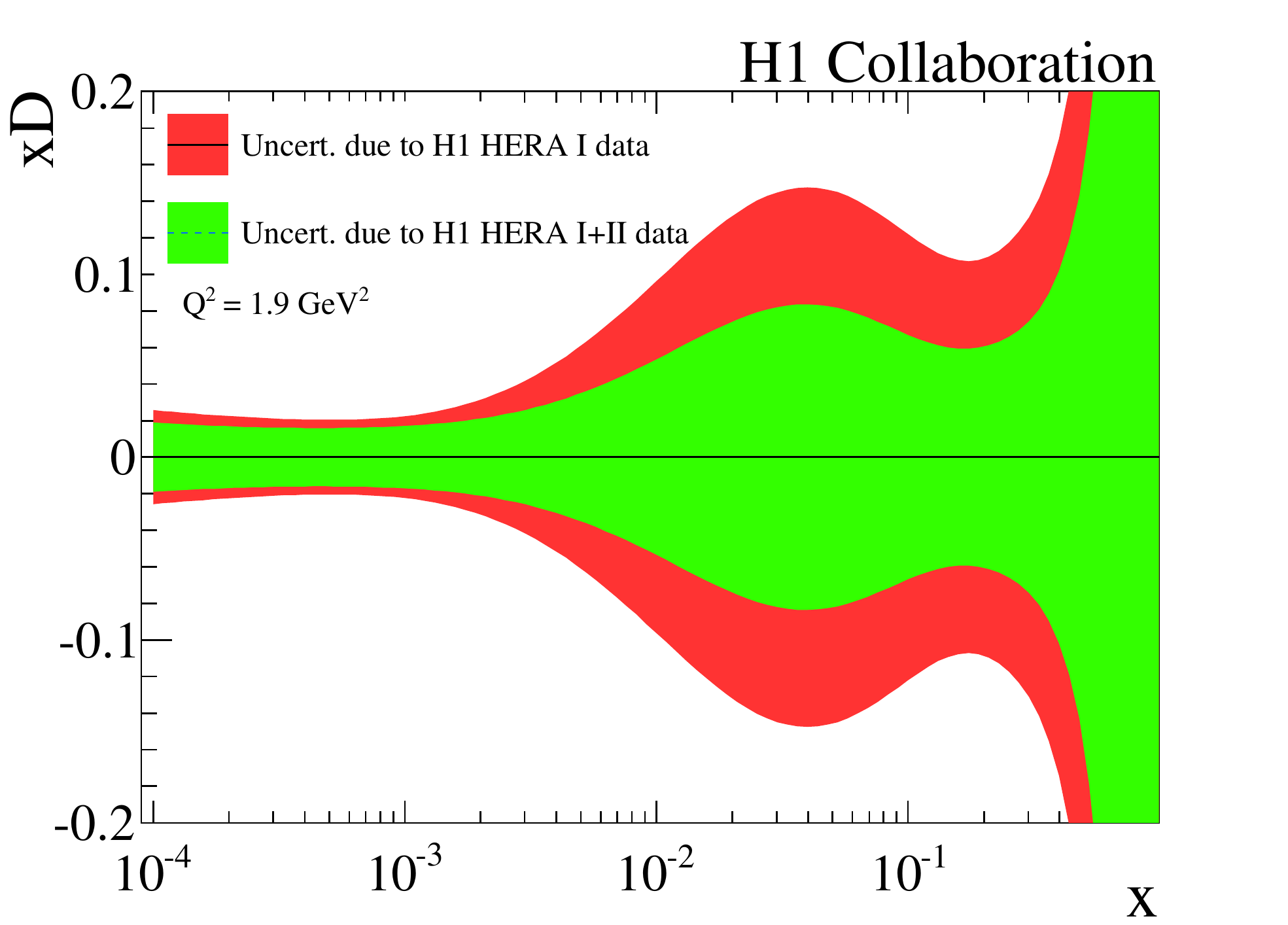}
\end{center}
\vspace{-5mm}
\caption{\underline{Left}: Parton distribution functions of H1PDF\,2012 at the evolved scale of $M^2_W$. The gluon and sea distributions are scaled by a factor 0.01. The uncertainties include the experimental uncertainties (inner), the model uncertainties (middle) and the parametrization variation (outer). All uncertainties are added in quadrature. \underline{Right}: Comparison of relative experimental uncertainties of $xD$ extracted from HERA-I (outer) vs.\ HERA\,I+II (inner) data sets under the same fit conditions to better assess the effect of the new high $Q^2$ measurements.}
\label{fig:pdfs}
\end{figure}

To assess the impact of the new H1 NC and CC cross sections at high $Q^2$ measured with the longitudinally polarised electron beams on the determination of PDFs, a new QCD analysis (H1PDF\,2012) has been performed. In addition to the new HERA\,II data, the previously published unpolarised HERA\,I data at high $Q^2$ and at low $Q^2$, as well as the H1 measurements at lower proton beam energies have been used. This analysis supersedes the previous H1PDF\,2009 fit. The resulting PDFs are shown in Fig.\ref{fig:pdfs}(left) and the impact of HERA\,II data on the PDF precision is shown as an example for the down quark density $xD$ in Fig.\ref{fig:pdfs}(right). 

\section{Summary}
%Using the low $E_p$ data taken at HERA-II in combination with the nominal $E_p$ data from both HERA-I and HERA-II, a direct $F_L$ measurement has been performed at an extended kinematic region $1.5\leq Q^2\leq 45\,{\rm GeV}^2$ and $2.7\times 10^{-5}<x<2\times 10^{-3}$ providing a direct test on the gluon density of the proton at lower $Q^2$ and $x$ than that of previous measurements. The integrated luminosity at HERA-II has been measured with elastic QED Compton events with atotal precision of 2.3\% better than the precision obtained with the Bethe-Heitler process for this data set. 
The NC and CC cross sections at high $Q^2$ with the nominal $E_p$ and all polarised $e^\pm$ data from HERA\,II have been measured. Both the statistical and systematic uncertainties of the measurements have been substantially reduced compared to previous HERA\,I H1 data. The NC lepton polarisation asymmetry $A^\pm$, sensitive to parity violation, has been determined separately for $e^+p$ and $e^-p$ scattering. The asymmetry is found to increase in magnitude with $Q^2$ in agreement with the expectation of the SM. The structure function $F_2^{\gamma Z}$ has been measured for the first time using the polarisation dependence of the $e^\pm p$ NC cross sections. At high $Q^2$ the structure function $xF_3^{\gamma Z}$ has been determined using unpolarised NC cross sections obtained from the complete HERA\,I and HERA\,II data sets. The polarization dependence of the CC total cross section for $Q^2>400\,{\rm GeV}^2$ and $y<0.9$ has been measured and compared to the unpolarised HERA\,I measurements. The data exhibit a linear scaling of the cross sections with $P_e$ which is positive for $e^+p$ and negative for $e^-p$ scattering. The data are consistent with the absence of right handed weak currents.  The new QCD analysis  shows a significant impact of the new data on the increased precision of the PDFs. These data are being combined with those from ZEUS and a new QCD fit HERAPDF2.0 is being performed. The combined data are expected to provide a better precision on PDFs in particular at medium and high $x$.

\end{document}